# Electrically-activated spin-controlled orbital angular momentum multiplexer

Gianluca Ruffato[1,2], Etienne Brasselet[3], Michele Massari[1,2], and Filippo Romanato[1,2,4]

1 Department of Physics and Astronomy 'G. Galilei', University of Padova, via Marzolo 8, 35131 Padova, Italy

2 LaNN, Laboratory for Nanofabrication of Nanodevices, EcamRicert, Corso Stati Uniti 4, 35127 Padova, Italy

3 University of Bordeaux, CNRS, LOMA, UMR 5798, Talence, France

4 CNR-INFM TASC IOM National Laboratory, S.S. 14 Km 163.5, 34012 Basovizza, Trieste, Italy

## ABSTRACT

We present and test the integration of a static orbital angular momentum mode multiplexer with a dynamical geometric-phase optical element enabling on-demand spin-controlled angular momentum multiplexing. A diffractive optics multiplexer fabricated with 3D high-resolution electron beam lithography performs a conformal mapping for the conversion from linear to azimuthal phase gradients. The latter is functionalized by a dynamic spin-orbit add-on that consists of a self-engineered electrically-activated liquid crystal optical vortex generator having large clear-aperture and high-resolution. By combining several functionalities based on the optical angular momentum of light in a compact manner, the proposed hybrid device could find applications in next-generation high-dimensional mode switchers and routers based on orbital angular momentum.

During the last decades, the manipulation of the phase and intensity distributions of light beams has attracted increasing attention for a wide range of applications that include particle trapping and manipulation [1], high-resolution optical imaging [2, 3], astronomical coronagraphy [4, 5], security [6, 7] and optical transmission [8]. In the telecom field, the ever-increasing need for information capacity and spectral efficiency of optical networks has eventually stimulated the use of the spatial degrees of freedom of light. Indeed, these allow to generate, propagate and detect a larger number of orthogonal optical states that could be used as distinct and independent information channels at the same frequency, in the so-called mode-division multiplexing (MDM) scheme [9]. In particular, modes carrying orbital angular momentum (OAM) have been demonstrated to provide a promising solution for both free-space [10] and optical fiber [11] propagation of light, either in the classical or in the quantum regimes [12]. Such modes are endowed with helical wavefronts associated with an optical phase singularity expressed by a phase term $exp(i\ell\varphi)$ ($\ell=0,\pm1,\pm2,…$), where $\varphi$ is the polar angle in a plane transverse to the propagation direction and $\ell$ is the amount of OAM per photon in units of $\hbar$ [13], and exhibit an axially-symmetric intensity distribution around a central dark point [14].

A pivotal part of an optical link based on OAM-MDM is the so-called multiplexer, which is the device generating a large set of collimated and coaxial orthogonal beams at the source. These optical states should be eventually sorted at the receiver according to their OAM modal content. This operation is ensured by a second device, the demultiplexer, that is typically a multiplexer working in reverse. To date, several solutions have been designed and experimentally tested [15], which cover a rather large range of complexity, efficiency and integration level [16]. Nowadays, it becomes essential to further develop novel architectures that can control and switch between distinct OAM modes dynamically [17] in order to fully exploit the extra degree of freedom provided by the OAM for both classical and quantum communications. Here, we propose to cascade two high-resolution flat-optics elements, namely a diffractive OAM multiplexer with a dynamical OAM shifter. This is done in a compact manner, without additional technological complexity and without clear-aperture drawbacks, by using a recently introduced self-engineered liquid crystal optical element [18]. This allows us to report experimentally on the demonstration of on-demand spin-controlled OAM multiplexing of light.

Among all the OAM-mode demultiplexing techniques, one of the most effective methods is the so-called transformation optics [19, 20]. It is widely used to sort the OAM, for example, in recent telecom experiments both in the classical [21] and quantum [22] regimes. This technique involves a unitary transformation converting the azimuthal phase gradients of OAM beams into linear phase gradients (i.e., tilted beams), which are then univocally mapped to spatial positions by means of optical Fourier transform using a spherical lens. This optical operation can be decomposed into two steps, each of these being ensured by a distinct element: the unwrapper (UW) and the phase-corrector (PC). The former performs a conformal mapping of a point ($x$, $y$) in the input plane to a point ($u$, $v$) in the output plane, where $v = a \arctan(y/x)$ and $u = -a \ln(r/b)$, being $r=(x^2+y^2)^{1/2}$, with $a$ and $b$ being design parameters. On the other hand, the latter corrects the resultant distorted phase by taking into account the optical path differences at each point, thus completing the conversion of the input azimuthal phase gradient into a linear one. The transmission phase masks associated to these two elements are respectively $exp(i\Omega_{UW})$ and $exp(i\Omega_{PC})$, where the phase functions are given by

$$\Omega_{UW}(x,y) = \frac{2\pi a}{\lambda f}\left[ y\arctan\left(\frac{y}{x}\right) - x\ln\left(\frac{r}{b}\right) + x \right] - \frac{2\pi}{\lambda}\frac{x^2+y^2}{2f}$$
$$\Omega_{PC}(u,v) = -\frac{2\pi ab}{\lambda f}\exp\left(-\frac{u}{a}\right)\cos\left(\frac{v}{a}\right) - \frac{2\pi}{\lambda}\frac{u^2+v^2}{2f}$$
(1)

Placing a lens in *f-f* configuration, the output truncated tilted wave is focused at a distance from the axis of the optical system that is proportional to the input OAM content, $\ell$, according to:

$$y_\ell = \frac{\lambda f}{2\pi a}\ell$$
(2)

When properly illuminated in the reverse order, these two elements behave as a multiplexer [23], thus mapping linear phase gradients into azimuthal ones. This was first implemented with spatial light modulators [19], and later by refractive optical components that bring higher efficiency [20]. In the quest for miniaturization, refractive 3D micro-optics have been realized recently, however for limited extent of OAM range sorting [24]. More recently, a flat-optics diffractive version has been proposed [25, 26] both for sorting and demultiplexing, further increasing the level of compactness and integration with respect to the refractive macroscopic counterpart. In the multiplexing stage, an initial reshaping of the input beam is required, by using for instance a cylindrical lens, in order to prepare the elongated spot to be wrapped by the first element [25].

In this work, the diffractive (de)multiplexers are fabricated with electron-beam lithography, by patterning a layer of poly(methyl methacrylate) (PMMA) resist (thickness of 2 μm, molecular weight of 950 kg/mol), spin-coated on a 1.1 mm thick ITO coated soda lime float glass substrate and pre-baked for 10 min at 180°C on a hot plate. The phase patterns are written with a JBX-6300FS JEOL EBL machine, 12 MHz, in high-resolution mode, generating at 100 KeV and 100 pA an electron-beam with a diameter of 2 nm, assuring a resolution down to 5 nm. Next, samples are developed for 1 min in a temperature-controlled bath (deionized water: isopropyl alcohol (IPA) 3:7) set on a magnetic stirrer at 1000 rpm. After development, the optical elements are gently rinsed in deionized water and blow-dried under nitrogen flux. At the experimental wavelength of the laser ($\lambda$ = 632.8 nm), the PMMA refractive index results $n_{PMMA}$ = 1.489, as measured by analysis with a spectroscopic ellipsometer (J.A. Woollam VASE, 0.3 nm spectral resolution, 0.005° angular resolution). A desired phase pattern $\Omega(x,y)$ is obtained by adjusting the depth $t(x, y)$ of the exposed zone for normal incidence in air following

$$t(x,y) = \frac{\lambda}{n_{PMMA}-1} \cdot \frac{2\pi - \Omega(x,y)}{2\pi} \tag{3}$$

The phase patterns are discretized into ~25 megapixels with 256 phase levels and with a single pixel area 0.312 x 0.312 μm² for a total size of 1.6 x 1.6 mm². From Eq.(3) the maximal depth of the surface relief pattern is 1289.0 nm, while the achieved thickness resolution is $\Delta t$ = 5.1 nm. Design parameters are $a$=220 μm, $b$=50 μm, $f$=9 mm. In addition, a tilt term is added to the phase-corrector in the sorter (spatial frequencies $\alpha$=$\beta$=0.1 μm$^{-1}$), in order to prevent the beam from overlapping with a possible zero-order term. The UW and PC elements are shown in Fig. 1(a) and 1(b), respectively.

The dynamical add-on is a liquid crystal (LC) geometric-phase optical element that plays the role of a $q$-plate [20] that is a space-variant half-wave plate whose optical axis orientation angle is given by $\psi$=$q\varphi$. Such an element behaves as a transmission phase mask associated with the polarization-dependent phase functions $\Omega_{LC}$= $2\sigma q\varphi$ on the circular polarization basis, where $\sigma$=±1 refers to the helicity of light. Here we implement a novel approach combining high-resolution of the phase pattern, large clear aperture and electrical reconfigurability, without need of any machining technique [18]. It consists in using a layer of nematic LC (mixture MLC-2048, Merck) with thickness $L = 20$ μm that is sandwiched between two glass substrates provided with transparent electrodes. Also, a thin layer of polyimide is deposited on the substrates to ensure perpendicular orientation of the LC molecules at both ends of the film. The LC is chosen to have a negative dielectric permittivity at the used quasi-static voltage frequency (here, $\varepsilon_a$=-3.4 at 100 kHz), which leads to the generation of randomly distributed microscopic self-engineered $q$-plates with $q$=±1 above a threshold voltage [27]. We get rid of this drawback by adding a static magnetic field from a ring-shaped macroscopic (centimetre size) permanent magnet [18], which pins a defect with $q$=+1 centered on the revolution axis of the magnet while preserving the inherent high-quality topological ordering of self-engineered approaches [27-30] as well as the electrical control features of liquid crystal $q$-plates [31]. Obtained macroscopic magneto-electric optical vortex phase mask observed between crossed linear polarizers is shown in Fig. 1(c) under white light illumination and a zoom of its central part is shown in Fig. 1(d) for the operating wavelength of present study ($\lambda$=633 nm). This solution provides a cheap, easy and robust method for the fabrication and integration of a $q$-plate optical component in existing devices.

The assembled set of three independent phase elements, where the liquid crystal element is packed at the unwrapper back-side with a thin film of immersion oil, is depicted in Fig. 1(e). Its behaviour is illustrated in Fig. 2 and can be summarized as follows. The LC has no detectable influence on the OAM content of the light coming out from the multiplexer that passes through in the absence of applied voltage, while it imparts to the transmitted light field an extra phase term $exp(\pm i2\varphi)$ for left-handed and right-handed circularly polarized light, respectively. In other words, it behaves as an electrically-activated spin-controlled OAM shifter providing a global OAM shift by an amount $\Delta\ell=\pm2$.

The experimental setup is sketched in Fig. 3. A Gaussian beam ($\lambda= 632.8$ nm, beam waist $w_0=240$ μm, power 0.8 mW) emitted by a HeNe laser source (HNR008R, Thorlabs) is reshaped by the cylindrical lens $L_1$ (focal length $f_1=2.54$ cm) and collimated by the lens $L_2$ ($f_2=7.5$ cm), in order to prepare an asymmetric input beam to be wrapped into a doughnut by the multiplexer, as sketched in Fig. 2. Both the laser and the reshaping lenses are mounted on the same stage and can be translated together using a micrometric translator (TADC-651, Optosigma) in the direction orthogonal to the propagation direction. The transmitted beam is Fourier-transformed with the lens $L_3$ ($f_3=20$ cm), which converts the axial displacement of the laser into an angular deviation for the beam that illuminates the first diffractive element (phase-corrector). Both the phase-corrector and the unwrapper are mounted on 6-axis kinematic mounts (K6XS, Thorlabs). A neodymium ring magnet is placed in contact with the LC plate. Its position can be adjusted with a sample holder with micrometric drives (ST1XY-S/M, Thorlabs), in order to put its center on the optical axis. The two electrodes of the LC plate are connected to a waveform generator (LeCroy LW120). The half-wave plate retardation fulfilment for the LC $q$-plate at the operating wavelength is ensured by adjusting the applied voltage. In practice, this is made by nulling the output power fraction of the co-circularly polarized component for a circularly polarized incident field. In present case, this corresponds to a square waveform voltage amplitude $U=3.21$ V at 100 kHz frequency. A beam-splitter is used to analyze the field profile, which is collected with a first CCD camera (DCC1545M,

Thorlabs). Then the beam illuminates the demultiplexing sequence. At last, the far-field is collected by a second CCD camera placed at the back-focal plane of the lens $L_4$ ($f_4$=10 cm).

A demonstration of OAM shifting is displayed in Fig. 4, where the spatio-temporal behaviour of the signal collected by $CCD_2$ is shown for an input left-handed circular polarization ($\sigma$=+1) over an on-and-off cycle for the electrically activated OAM shifter. As expected, the output spot shifts from the position that corresponds to $\ell$=0 to the one that corresponds to $\ell$=+2 after the LC plate is turned on and the process is reversed when the LC is turned off. The on/off characteristic times are respectively given by the liquid crystal electrical reorientation time and the elastic relaxation time. The latter merely depends on the thickness of the LC layer and scales as $1/L^2$ while the former depends on the initial and final reorientation states and is rather large when the off-state corresponds to the non-reoriented state. The switch-on time could be straightforwardly improved by working between the half-waveplate and full-wave retardations conditions.

The generation and detection of OAM modes in the range from $\ell$=-6 to $\ell$=+6 has been studied for the three cases illustrated on Fig. 2. Namely, LC plate off (in that case the results are independent on the incident polarization state) and LC plate on with either right- or left-handed incident circular polarization states. The data collected on $CCD_1$ (far-field of OAM modes) and $CCD_2$ (OAM channels) are illustrated qualitatively in Fig. 5 for three different incident OAM channels ($\ell$ = -2, 0 and +3). As expected, the output mode exhibits a shift of two units of OAM depending on its polarisation state $\sigma$, and, consequently, the far-field spot appears at a new shifted position after demultiplexing:

$$y_\ell^{on,\sigma} = y_\ell^{off} + \Delta y^{on,\sigma} = \frac{\lambda f}{2\pi a}(\ell + 2\sigma) \qquad (4)$$

In Fig. 6, the positions of the far-field spots are quantitatively investigated in the range of incident OAM states, represented here as a function of the input laser lateral shift. The linear response of the system is confirmed for both configurations with LC on and off, and the measured shifts induced by the OAM shifter fits to a variation that corresponds to $\Delta\ell=\pm 2$ within the experimental errors.

To conclude, we presented and tested a novel and compact optical device integrating an OAM multiplexer with an electrically-activated spin-controlled liquid crystal OAM shifter. This optical element combines distinct optical operations on the OAM content of a light field with a high miniaturization level. At the same time, the presented architecture is compatible with mass-production, since EBL masters can be replicated with cheap and fast soft-lithography techniques, while no patterning is required for the generation of the $q$-plate, which originates from a self-engineering process arising in an initially homogeneous liquid crystal slab under a magneto-electric external stimulus. This device can find applications in polarization-dependent OAM-mode multiplexers and sorters, and can represent a key-element in the design of optical architectures for OAM-mode switching and routing [32-34].

**Acknowledgments**

We thank Mireille Quémener and Tigran Galstian from Laval University (Quebec, Canada) for their kind preparation of the liquid crystal sample.

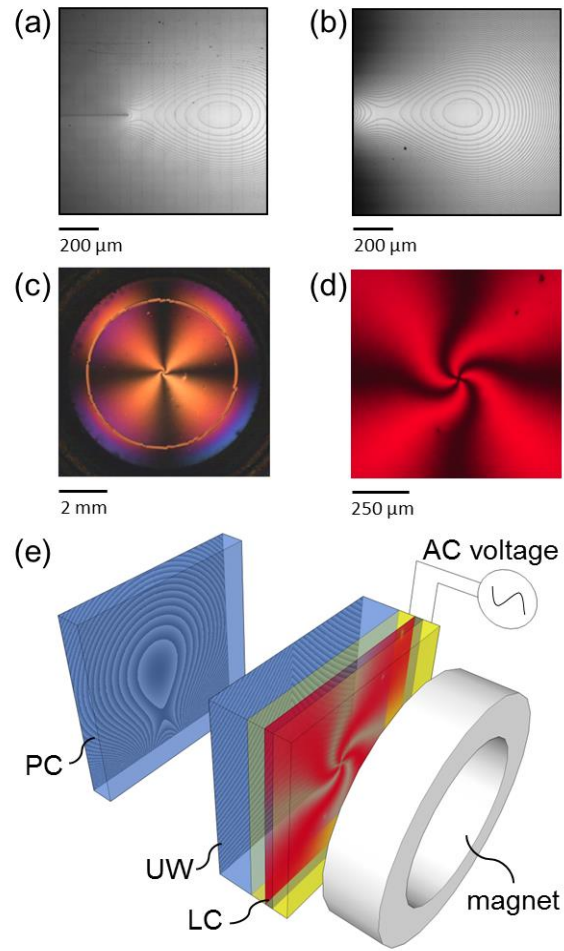

**Figure 1.** Optical microscope inspections of unwrapper (a) and phase-corrector (b) diffractive optics. (c) Obtained macroscopic magneto-electric optical vortex phase mask observed between crossed linear polarizers under white light illumination. (d) Zoom of its central part for the operating wavelength ($\lambda$=633 nm). (e) Scheme of the assembled set of three independent phase elements constituting the electrically spin-controlled multiplexer: phase-corrector (PC), unwrapper (UW), magneto-electric liquid-crystal plate (LC) (not to scale).

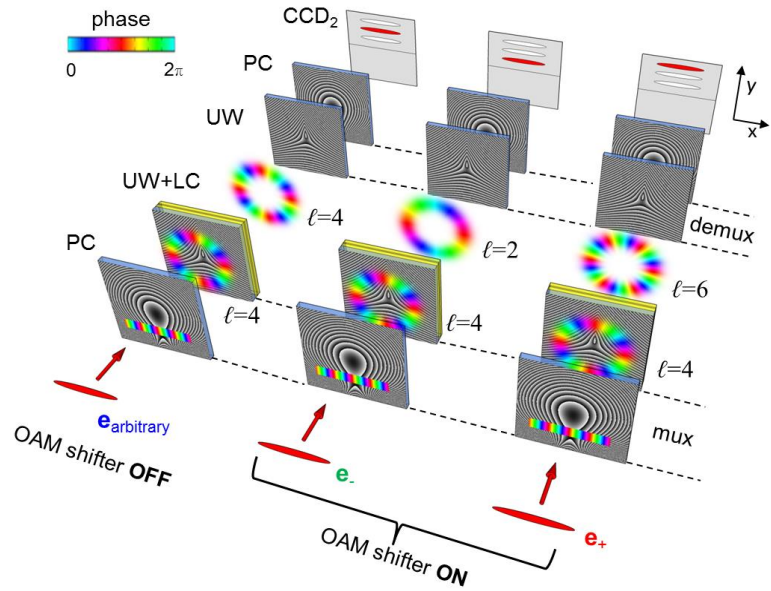

**Figure 2.** Pictorial representation of the OAM-shifter working principle for input mode corresponding to $\ell=+4$. A sequence of diffractive optics performing *log-pol* optical transformation, i.e. unwrapper (UW) and phase-corrector (PC) is exploited for demultiplexing and multiplexing (in reverse order). The multiplexing sequence PC-UW converts the input linear gradient into an output azimuthal gradient, generating an OAM beam. The liquid-crystal (LC) plate, matched to the second optics (UW), allows shifting the OAM content of the output beam of $\Delta\ell=\pm2$, depending on the handedness of the input circular polarization ($\sigma=\pm1$). Finally, the sequence UW-PC performs the demultiplexing of the OAM-shifted beam.

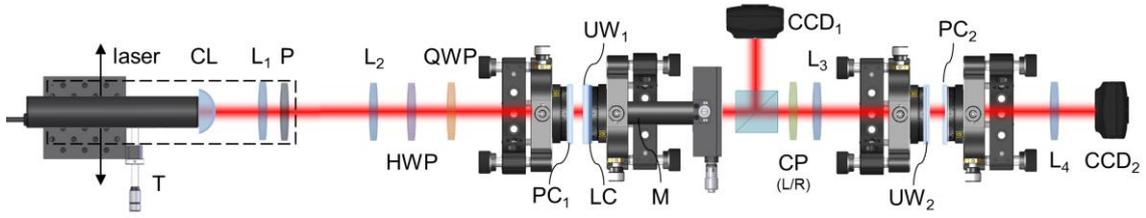

**Figure 3.** Scheme of the optical setup for OAM-shift analysis. The Gaussian beam emitted from the HeNe laser illuminates, in sequence, a cylindrical lens (CL) and a first lens $L_1$ for beam reshaping and collimation. The input system can move perpendicularly to the propagation direction with a micrometric translator (T). The lateral shift is Fourier-transformed into an angular tilt by a lens ($L_2$) and the beam illuminates the sequence of optical elements $PC_1$-$UW_1$ for multiplexing. The liquid-crystal plate (LC) is matched to the $UW_1$ back-side with a thin layer of immersion oil. A distinct sample-holder with *xy* micrometric translators controls the position of the magnet (M) on the LC surface. The input polarization state is controlled with the sequence of linear polarizer (P), half-wave plate (HWP) and quart-wave plate (QWP). Next, the generated optical vortex is split with a 50:50 beam-splitter (BS) for beam analysis with a first camera ($CCD_1$). The second part of the beam in focused ($L_3$) on the demultiplexing sequence $UW_2$-$PC_2$ and it is finally Fourier-transformed by a lens ($L_4$) and collected ($CCD_2$). Left/right-handed circular polarizers (L/R CP) are mounted after the muxer during the tuning of the LC voltage amplitude in order to check the total polarization conversion.

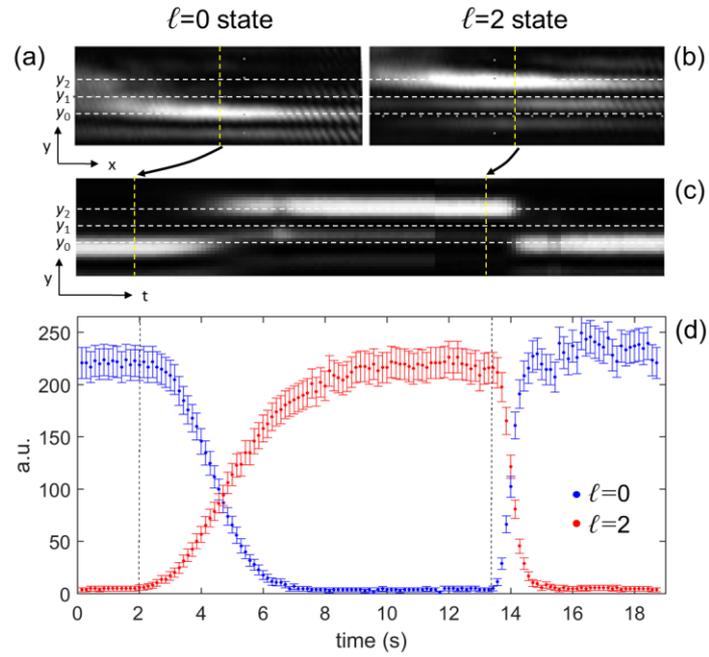

**Figure 4.** Spatio-temporal behaviour of the signal collected for an input left-handed circular polarization ($\sigma=+1$) over an on-and-off cycle for the electrically activated OAM shifter. Far-field collected before (a) and after (b) the OAM-shifter activation. (c) Evolution over time of the intensity collected along a cross-section and (d) in correspondence of two points at the positions $y_0$ and $y_2$. Voltage amplitude is adjusted to fulfil the half-wave retardance condition for the working wavelength $\lambda=632.8$ nm at $f=100$ kHz, namely $U=3.21$ V (square waveform).

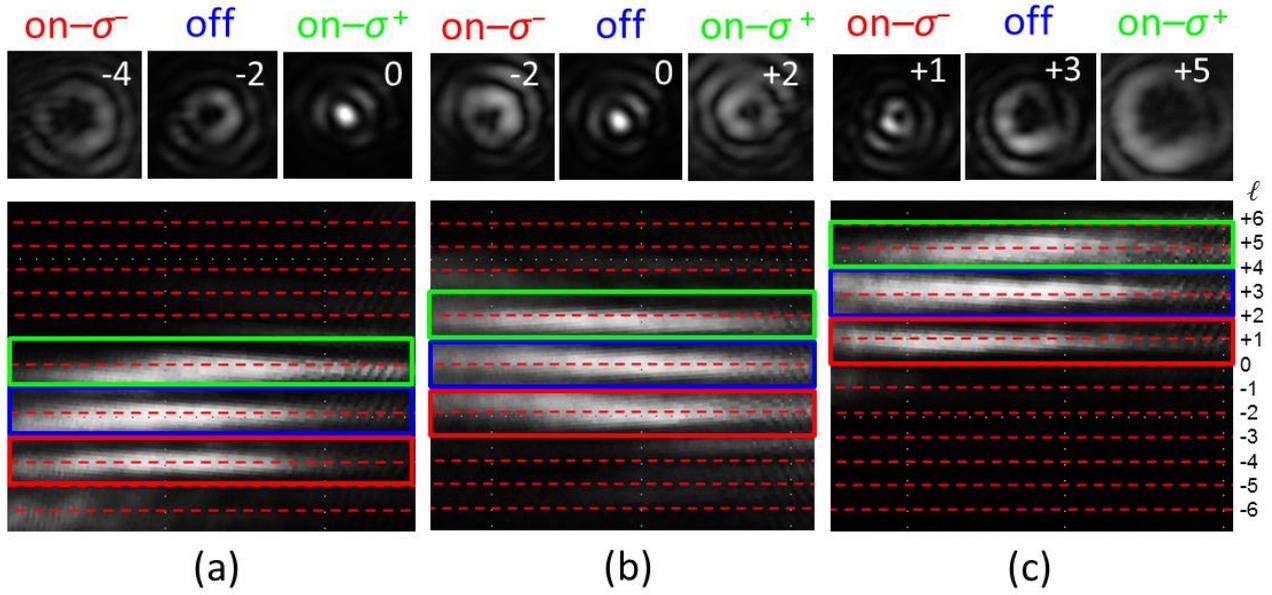

**Figure 5.** Generated OAM beams (collected on $CCD_1$, see Fig. 3) and corresponding far-field spots (on $CCD_2$), detected when the LC plate is off (blue) and on, for left-handed (green) and right-handed (red) circular polarizations, for three different positions of the input laser, corresponding to $\ell=-2$ (a), $\ell=0$ (b) and $\ell=+3$ (c) when the LC plate is turned off.

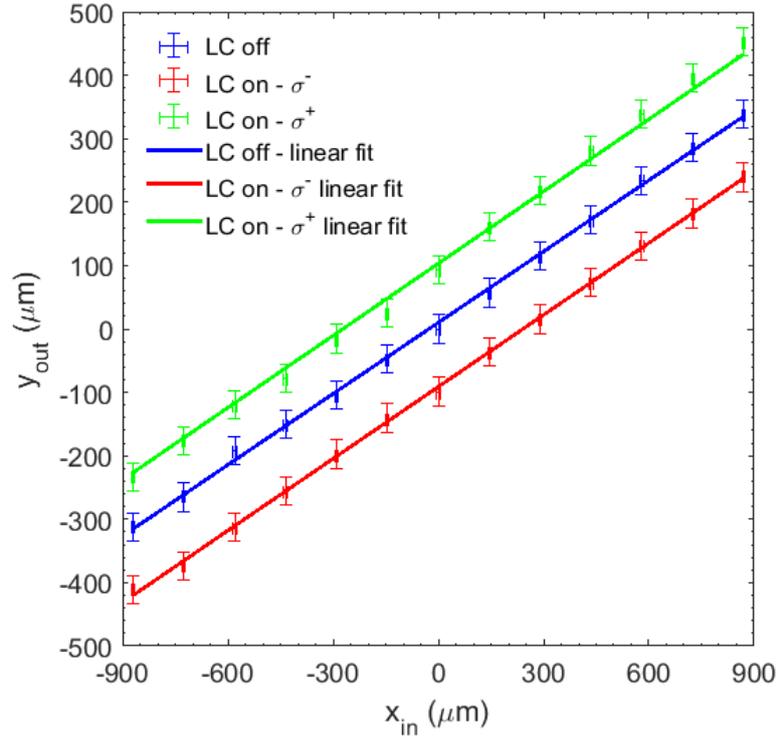

**Figure 6.** Positions of the output spots in far-field as a function of the input laser lateral shift when the LC plate is off (blue) and on, for left-handed (green) and right-handed (red) circular polarizations. Linear fit with $y=a \cdot x+b$: $(a, b)_{off}$ = (0.374±0.004, 7.0±1.9), $(a, b)_{on,+}$ = (0.380±0.005, 103.2±2.4), $(a, b)_{on,-}$ = (0.378±0.003, -91.2±1.4). $\Delta b^+$ = (98.2±4.3) µm, $\Delta b^-$ = (-96.5±3.3) µm. Theoretical shift: $\Delta y(\Delta \ell=2)$ = 92.0 µm.